\documentstyle[11pt]{article}
\def\deg    {$^\circ$}

\def\eg   {{\it e.g.\ }}

\begin{document}
\twocolumn

\section*{Tunable Imaging Filters}
\author{Joss Bland-Hawthorn}

While tunable filters are a recent development in night time astronomy,
they have long been used in other physical sciences, \eg solar physics,
remote sensing and underwater communications.  With their ability to tune 
precisely to a given wavelength using a bandpass optimized for the 
experiment, tunable filters are already producing some of the deepest 
narrowband images to date of astrophysical sources.  Furthermore, some 
classes of tunable filters can be used in fast telescope beams and 
therefore allow for narrowband imaging over angular fields of more than a 
degree over the sky.

\section*{The physics of tunable imaging}

A rich variety of physical phenomena can isolate a finite spectral 
band: absorption, scattering, diffraction, evanescence,
birefringence, acousto-optics, single layer and multi-layer 
interference, multi-path interferometry, polarizability,
and so on.  Most of the available optical filter technologies are 
listed below:

\medskip
{\sl

linear / circular variable filter

multi-layer dielectric filter

Fabry-Perot interferometer

Michelson interferometer

acousto-optic filter

solid etalon filter

solid Michelson filter

generalized resonant grating filter 

sub-lambda evanescent grating filter

volume phase holographic grating filter

Lyot-\"{O}hman filter

generalized Lyot filter

generalized \u{S}olc filter

liquid crystal filter 
}

\medskip\noindent
There is a bewildering array of future possibilities, including tunable 
crystal lattice structures, and many possible variants on the above
technologies. Here, we focus on a few key technologies.

\section*{The ideal device}

The ideal filter is an imaging device which can isolate
an arbitrary spectral band $\delta\lambda$ at an arbitrary wavelength
$\lambda$ over a broad, continuous spectral range, preferably with 
a response function which is identical in form at all wavelengths.

The tilting interference filter is much the worst form of tunable filter.
The spectral range covered is almost negligible, and the filter profile 
varies with tilt angle.  Some devices (\eg Lyot, linear variable filter)
work only at a fixed resolving power but have the ability to tune 
over a wide spectral window.  Better devices allow for a wide 
selection of resolving powers over a wide range of wavelengths. 

The different techniques rely ultimately on the interference of beams that 
traverse different optical paths to form a signal. The technologies which 
come closest to the ideal tunable filter are the air-gap Fabry-Perot and
Michelson (Fourier Transform) interferometers. To understand why, we
highlight the Taurus Tunable Filter (TTF) which was the first general 
purpose device for night-time astronomy \break
(see http://www.aao.gov.au/local/www/jbh/ttf).
This is a Fabry-Perot filter
where interference is formed between two highly reflective, moving plates.
To be a useful filter, not only must the plates move through a large 
physical range, but they must start at separations of only a few 
wavelengths, as we show.

The condition for photons with wavelength $\lambda$ to pass through
the filter is (see Fig. 1)
\begin{equation}
m\lambda = 2\mu\ell\ \cos\theta
\label{intf}
\end{equation}
from which it follows that
\begin{equation}
{{d\cal R}\over{\cal R}}\ = \ {{dm}\over{m}}\ = \ {{d\ell}\over{\ell}} .
\end{equation}
For an order of interference $m$, the resolving power is 
$R = m {\cal N}$ where ${\cal N}$ is the instrumental finesse.
The finesse is determined by the coating reflectivity and is
essentially the number of recombining beams. For the TTF,
the plates can be scanned over the range $\ell=1.5-15\mu$m, and the
orders of interference span the range $m=4-40$, such that the 
available resolving powers are $R=100-1000$.

The sharp core of almost all tunable-filter transmission profiles is
not ideal. Even a small amount of flatness at peak 
transmission can ensure that we avoid narrowly missing most of the 
spectral line signal from a source.  In theory, all band-limited functions 
can be squared off, but in practice this is difficult for all but two
devices. Since the Michelson interferometer (filter) obtains its data in the 
frequency domain, the profile can be partially squared off at the data 
reduction stage through a suitable choice of convolving function. For the 
\u{S}olc filter, the instrumental profile can be modified through the 
use of partial polarizers and birefringent retarder elements.

\section*{Summary of filter technologies}

Here, we provide a brief outline of some of the key technologies.

\subsection*{Monolithic filters}

\noindent{\it Interference filter:}
The principle relies on a dielectric spacer sandwiched between two
transmitting layers (single cavity). The substrates are commonly fused
silica in the ultraviolet, glass or quartz in the optical, and 
water-free silica in the infrared. Between the spacer and the glass,
surface coatings are deposited by evaporation which partly transmit
and reflect an incident ray. Each internally reflected ray shares 
a fixed phase relationship to all the other internally reflected
rays. For constructive interference, for a wavelength $\lambda$ to
be transmitted, it must satisfy eqn.~1
where $\theta=\theta_R$ is the refracted angle within the optical spacer,
and the optical gap is the product of the thickness $\ell$ and 
refractive index $\mu$ of the spacer. The construction of these filters 
has undergone a revolution through the use of dielectric, multi-layer 
thin film coatings, and a proper description
is more involved. All such filters can be tuned through a small
wavelength interval ($\delta\lambda/\lambda = -\theta_R^2/2\mu^2$),
which amounts to no more than $2\%\lambda$ in practice.  Suffice it to 
say, interference filters make for poor tunable devices.

\subsection*{Gap-scanning filters}

\smallskip\noindent{\it Fabry-Perot filter:}
The air-gap etalon, or Fabry-Perot filter, was introduced in the previous 
section. 
The etalon comprises two plates of glass kept parallel over a small separation 
where the inner surfaces are mirrors coated with high reflectivity $\Re$. The 
transmission of the etalon to a monochromatic source $\lambda$ is given by the 
Airy function
\begin{equation}
{\cal A} = \left( 1+{{4\Re}\over{(1-\Re)^2}}\sin^2 (2\pi\mu \ell\cos\theta/\lambda) \right)^{-1}
\end{equation}
where $\theta$ is the off-axis angle of the incoming ray and $\mu l$ is the 
optical gap.  The condition for peaks in transmission is given in eqn.~1.
Note that
$\lambda$ can be scanned physically in a given order by changing $\theta$
(tilt scanning), $\mu$ (pressure scanning), or $l$ (gap scanning). 
Both tilt and pressure scanning suffer from serious drawbacks which limit 
their dynamic range. With the advent of servo-controlled, capacitance 
micrometry, the performance of gap scanning etalons surpasses other 
techniques. These employ piezo-electric transducers that undergo dimensional 
changes in an applied electric field, or develop an electric field when 
strained mechanically. Queensgate Instruments, Ltd. have shown that it is 
possible to maintain plate parallelism to an accuracy of $\lambda/200$ while 
continuously scanning over several adjacent orders.

Fabry-Perot filters have been made with 15~cm apertures and physical
scan ranges up to 3~cm.  The etalon is ultimately limited 
by the finite coating thickness between the mirrors, so it really only 
achieves the lowest interference orders ($m<5$) at infrared wavelengths.

\smallskip\noindent{\it Solid etalon filter:}
These are single cavity Fabry-Perot devices with a transparent 
piezo-electric spacer, \eg lithium niobate. The thickness and, to a lesser
extent, refractive index can be modified by a voltage applied to both
faces. For low voltage systems, tilt and temperature can be used to 
fine-tune the bandpass. High quality spacers with thicknesses less than
a few hundred microns are difficult to manufacture, so that etalon filters
are normally operated at high orders of interference. The largest devices
we have seen are 5~cm in clear aperture.

\smallskip\noindent{\it Michelson filter:}
In the Fourier Transform or Michelson filter, the collimated beam is split 
into two paths at the front surface of the beam-splitter. The separate
beams then 
undergo different path lengths by reflections off separate mirrors before 
being imaged by the camera lens at the detector.  The device shown in Fig.~ 2 
uses only 50\% of the available light. As Maillard has demonstrated at the 
Canada France Hawaii Telescope, it is possible to recover this light but the 
layout is more involved. 

The output signal is a function of path difference between the mirrors. At 
zero path difference (or arm displacement), the waves for all frequencies 
interact coherently.  As the movable mirror is scanned, each input wavelength 
generates a series of transmission maxima.  Commercially available devices 
usually allow the mirror to be scanned continuously at constant speed, or to 
be stepped at equal increments. At a sufficiently large arm displacement, the 
beams lose their mutual coherence.

The filter is scanned from zero path length ($x=y=0$) to a maximum path length 
$y=L$ set by twice the maximum mirror spacing ($x=L/2$). The superposition of 
two coherent beams with amplitude $b_1$ and $b_2$ in complex notation is 
$b_1 + b_2 e^{i 2\pi\nu y}$ where $y$ is the total path difference and $\nu$ 
is the wavenumber. If the light rays have the same intensity, the combined 
intensity is $4 b^2 \cos^2 \pi\nu y$, where $b = b_1 = b_2$. The combined 
beams generate a series of intensity fringes at the detector.  If it was 
possible to scan over an infinite mirror spacing at infinitesimally small 
spacings of the mirror, the superposition would be represented by an ideal 
Fourier Transform pair, such that 
\begin{eqnarray}
b(y)   &=& \int^{\infty}_{-\infty} B(\nu) (1+\cos 2\pi\nu y)\ d\nu \\
B(\nu) &=& \int^{\infty}_{-\infty} b(y) (1+\cos 2\pi\nu y)\ dy
\end{eqnarray}
where $b(y)$ is the output signal as a function of pathlength $y$ and $B(\nu)$ 
is the spectrum we wish to determine. $B(\nu)$ and $b(y)$ are both undefined 
for $\nu < 0$ and $y<0$: we include the negative limits for convenience. 
Note that
\begin{eqnarray}
b(y)-{{1}\over{2}}b(0) &=& \int^{\infty}_{-\infty} B(\nu) \cos 2\pi\nu y\ d\nu \\
B(\nu) &=& \int^{\infty}_{-\infty} [b(y)-{{1}\over{2}}b(0)] \cos 2\pi\nu y\ dy
\end{eqnarray}
The quantity $b(y)-{{1}\over{2}}b(0)$ is usually referred to as the 
interferogram although this term is sometimes used for $b(y)$. The 
spectrum $B(\nu)$ is normally computed using widely available Fast Fourier 
Transform methods.  The construction of a Michelson filter is a major
optomechanical challenge.  The ideal Fourier Transform pair is never 
realized in practice.  However, the Michelson filter probably comes 
closest to achieving the goal of an ideal tunable filter.

The Michelson does not suffer the coating thickness problems of the 
Fabry-Perot filter, and therefore reaches the lowest orders even 
at optical wavelengths.

\subsection*{Grating filters}

\smallskip\noindent{\it Resonant grating filter:}
These novel filters are inspired by the diffractive colours in many
insects, and constitute dielectric gratings with three-dimensional,
sub-micron microstructure.  The zeroth order 
reflection exhibits a broad to intermediate bandwidth (${\cal R} \sim\ 20$),
is highly polarized and maintains useful efficiency over a $\pm 30$\deg\
tilt (or rotation) range. Gale (1998) presents one grating design that 
produces a roughly self-similar bandpass from 450 to 850~nm over this tilt 
range. Grating filters, and their close relatives, evanescent gratings,
show great promise but most have yet to leave the drawing board. 
However, since much of the research is driven by bank note security, 
we anticipate rapid progress. Volume phase holographic gratings $-$ in 
reflection $-$ can produce a highly efficient grating filter through Bragg 
diffraction.

\smallskip\noindent{\it Acousto-optic filter {\rm (AOTF)}:}
These are electronically tunable filters that make use of acousto-optic
(either collinear, or more usefully, non-collinear) diffraction in an 
optically anisotropic medium. 
AOTFs are formed by bonding piezo-electric
transducers such as lithium niobate to an anisotropic birefringent medium.
The medium has traditionally been a crystal, but polymers have been 
developed recently with variable and controllable birefringence. When
the transducers are excited to 10-250 MHz (radio) frequencies, the 
ultrasonic waves vibrate the crystal lattice to form a moving phase
pattern that acts as a diffraction grating. The diffraction angle
(and therefore wavelength) can be tuned by changing the radio frequency.  
These devices are often water cooled to assist the thermal dissipation,
although this is less important in the UV where AOTFs are particularly
useful. The largest devices are 2.5~cm in diameter since it proves to be 
difficult to maintain a uniform acoustic standing wave over larger areas.
An additional problem is the 15$\mu$m structure in the LiNO$_2$ crystal
which is not always optimal for good image quality. But the acceptance 
angle of the AOTF is generally larger than the Fabry-Perot.

\subsection*{Birefringent Filters}

The underlying principle of the birefringent filter is that light originating 
in a single polarization state can be made to interfere with itself. 
An optically anisotropic, birefringent medium 
can be used to produce a relative delay between ordinary and extraordinary 
rays aligned along the fast and slow axes of the crystal.  
(A birefringent medium has two different refractive indices, depending on the
plane of light propagation through the medium.) Title and collaborators have 
discussed at length the relative merits of different types of birefringent 
filters.  The filters are characterised by a series of perfect polarizers
(Lyot filter), partial polarizers, or only an entrance and an exit polarizer 
(\u{S}olc filter).
The highly anisotropic off-axis behaviour of uniaxial crystals give 
birefringent filters a major advantage. Their solid acceptance angle is one 
to two orders of magnitude larger than is possible with interference filters 
although this is partly offset by half the light being lost at the entrance 
polarizer. 

\smallskip\noindent{\it Lyot filter:} This is conceptually the easiest to 
understand and forms the basis for many variants. The entrance 
polarizer is oriented 45$^{\circ}$ to the fast and slow axes so that the 
linearly polarized, ordinary and extraordinary rays have equal intensity. The 
time delay through a crystal of thickness $d$ of one ray with respect to the 
other is simply $d\ \Delta\mu / c$ where $\Delta \mu$ is the difference in 
refractive index between the fast and slow axes.  The combined beam emerging 
from the exit polarizer shows intensity variations described by 
$I^2\cos(2\pi d\ \Delta\mu/\lambda)$ where $I$ is the wave amplitude. As 
originally illustrated by Lyot (see Fig.~3), we can isolate an arbitrarily 
narrow spectral 
band-pass by placing a number of birefringent crystals in sequence where each 
element is half the thickness of the preceding crystal. This also requires the 
use of a polarizer between each crystal so that the exit polarizer for any 
element serves as the entrance polarizer for the next.  The resolution of the 
instrument is dictated by the thickness of the thinnest element.
With quarter-wave plates placed between each of the retarder elements, 
$\lambda$ can be tuned over a wide spectral range by rotating the crystal 
elements. But to retain the transmissions in phase requires that each crystal 
element be rotated about the optical axis by half the angle of the preceding 
thicker crystal. 

Woodgate (NASA Goddard Space Flight Center) has made a Lyot filter utilising 
eight quartz retarders 
with a 10~cm entrance window. The retarders, each of which are sandwiched with 
half-wave and quarter-wave plates in addition to the polarizers, are rotated 
independently with stepping motors under computer control.  They achieve a 
bandpass of 0.4$-$0.8~nm tuneable over half the optical wavelength range 
(350$-$700~nm).  

\smallskip\noindent{\it \u{S}olc\footnote{The proper Czech pronunciation is 
`Sholtz'.} filter:} These highly non-intuitive filters use only two 
polarizers and a chain of identical retarders with varying position angles
(Evans 1958). There are folded (zigzag) and fanned designs with the former 
having the better performance. Title has made a tunable \u{S}olc 
filter with 7~cm clear aperture.  It has the extraordinary capability of
tuning the spectral profile:  an $n$-element \u{S}olc filter can have a 
profile that is determined by $n$ Fourier coefficients.  The same can be 
achieved with polarizing filters by proper choice of crystal lengths.

\smallskip\noindent{\it Liquid crystal filter {\rm (LCTF)}:}
These are rapid switching, electronically tuned devices which employ either 
ferroelectric or nematic liquid crystals (LC). The more commonly used nematic 
LCTF (Morris, Hoyt \& Treado 1994)
comprises a series of liquid crystal elements whose thicknesses are
cascaded in the same way as the Lyot filter. However, the tuning is
achieved by electronically rotating the crystal axes of the LC waveplate.
When no voltage is applied, the retardance is at a 
maximum; at large applied voltages, the retardance reaches a minimum.
The retardance can be tuned continuously to allow the wavelength to
be tuned. 

Liquid crystal filters are now commercially available from Cambridge 
Research \& Instrumentation, Inc. The biggest device we
have seen has a clear aperture of 4~cm, requires about 5~V to scan a single
order of interference, and appears to have good image quality. The tunable
band for a single stage device is about ${\cal R} \sim\ 5$ but can be tuned over 
the optical window.  The peak transmissions are 30\% or less.


\section*{Differential imaging}

There are many important reasons for pursuing tunable filter imaging,
as demonstrated by the TTF since the mid 1990s at the Anglo-Australian 
3.9m and William Herschel 4.2m telescopes. 
Conventional imaging has a major limitation in that images are taken 
sequentially which is not ideal at even the best sites. Even small
detector, instrument or atmospheric variations lead to systematic error
between images.  A better approach is a multi-band camera so that 
different bands are observed in parallel. However, even this is not 
ideal in that the optical path is different for each filter.
A tunable filter provides a powerful alternative
since band switching can be linked to charge shuffling with the CCD.
This is a truly differential technique 
which leads to much smaller systematic errors than is possible with 
conventional imaging.

\section*{Wide-field imaging}

Few telescopes offer more than a wide-field imager at prime 
focus simply because it is very difficult to exploit the fast beam 
spectroscopically. Spectral passbands are degraded in converging beams 
which is unfortunate as the widest fields are achieved at the fastest 
f/ratios.  The wide-field expanded Lyot filter is (almost) the last 
word in exploiting the widest possible field with a given 
telescope. Remarkably, beams as fast as f/2 can be compensated
with crossed birefringent elements, in concert with half-wave plates, 
such that even a constant sub-Angstrom bandpass is possible over a 
degree-sized field.  This opens up many new astronomical programs,
\eg Lyot filters can be vastly more 
efficient for redshift-targetted surveys of galaxies (\eg high redshift 
clusters) compared with existing multi-aperture spectrographs. 
To our knowledge, a wide-angle \u{S}olc filter using half-wave plates
has not been attempted although it is entirely feasible.

\section*{Bibliography \& Further Reading}

\medskip\noindent
Bell R J 1972 Introductory Fourier Transform Spectroscopy (San Diego: Academic 
Press)

\medskip\noindent
Evans J W 1958 The \u{S}olc birefringent filter {\it J. Opt. Soc. Amer.} 
{\bf 48} 142

\medskip\noindent
Gale M T 1998 In Optical document security, R L van Renesse ed. (Artech House) 
ch 12

\medskip\noindent
Morris H R, Hoyt C C and Treado P J 1994 Imaging spectrometers for fluorescent 
and Raman microscopy: acousto-optic and liquid crystal tunable filters 
{\it Applied Spectroscopy} {\bf 48} 857--866

\medskip\noindent
Title A M and Rosenberg W J 1981 Tunable birefringent filters {\it Opt. Eng.} {\bf 20} 815--823

\medskip\noindent
Vaughan J M 1989 The Fabry-Perot interferometer: history, theory, practice and
applications (Bristol: Adam Hilger)

\vskip 2cm
\section*{Figure captions}

\medskip\noindent {\bf Fig. 1.} 
(i) Interference filter: the internal structure is not shown. (ii) Fabry-Perot 
etalon. (iii) Wedged Fabry-Perot etalon which avoids the problem of the 
plates behaving like interference filters.

\medskip\noindent {\bf Fig. 2.} 
Schematic of a two-beam Michelson (Fourier Transform) interferometer.

\medskip\noindent {\bf Fig. 3.} 
The transmission profile of a simple Lyot cascade with (top to bottom)
1, 2, 3, 4 and 5 stages. The thinnest element has retardance R and the 
following elements have thicknesses which are multiples of this element. 
The polaroids P are aligned with each other and oriented at 45\deg\ to the 
retarders.

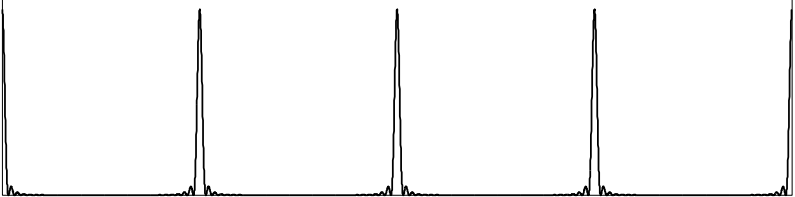
\end{document}